# Identifying the Alloy Structures of Germanene Grown on Al(111) Surface

Feini Yan,[a] Shaogang Xu,[a] Chao He,[a] Changchun He,[a] Changming Zhao,[a] and Hu Xu[a,*]

[a]Department of Physics, Southern University of Science and Technology, Shenzhen518055, China

**ABSTRACT**

**While the growth of germanene has been claimed on many substrates, the exact crystal structures remain controversial. Here, we systematically explore the possible structures formed by Ge deposition onto Al(111) surface by combining density-functional theory (DFT) and global optimization algorithm. We show that, by high level random-phase approximation (RPA) calculations, the formation of germanene on Al(111) is energetically unfavorable with positive formation energy. The two experimental phases are identified as honeycomb alloys $Al_3Ge_3$/Al(111)($\sqrt{7}\times\sqrt{7}$) and $Al_3Ge_4$/Al(111)($3\times3$), by combining ab initio evolutionary simulations, RPA calculations, and available experimental data from scanning tunneling microscopy (STM) and low-energy electron diffraction (LEED). $Al_3Ge_4$/Al(111)(3×3) is an interesting structure with a vacancy in the substrate, which accounts for the dark clover pattern in the experimental STM image. Our results clarify the structural controversy of the Ge/Al(111) system and indicate the fabrication of germanene may remain challenging.**

## Introduction

Germanene is an attractive two-dimensional material, as it is predicted to could exist in a freestanding honeycomb configuration[1-2] with extraordinary physical and chemical properties, such as quantum spin Hall effect (QSHE)[3], massless Dirac fermion system[2], and tunable gap via an external electric field[4]. Due to the lack of a layered bulk phase, the formation of germanene needs a substrate. Many substrates have been attempted to grow germanene, for example, the most commonly used metal substrates (Au[5], Ag[6-7], Al[8], Pt[9]), the semiconductor substrate $MoS_2$[10], and highly oriented pyrolytic graphite (HOPG)[11]. However, many results have been questioned lately, for instance, the structure on HOPG has been attributed to a charge density wave[12] and the structures on some metal substrates (Au[13-15], Ag[16-17], and Al[18]) surfaces have been considered to be surface alloys. In particular, to date, many of these structures are still under debate and no consensus has been reached. Since crystal structure is the basis for exploring physical and chemical properties, this structural uncertainty hinders a deep understanding of these materials. Therefore, it is of paramount importance to identify the exact crystal structure at the level of monolayers of atoms.

Among the various substrates, Al(111) surface is one of the most widely studied. The formation of germanene on this surface is first reported by Delivaz et al. in 2015[8], and then attracts great interest[19-30]. Two ordered phases have been observed experimentally, corresponding to the $\sqrt{7}\times\sqrt{7}$ reconstruction[19-21, 25] and $3\times3$ reconstruction[8, 19-24] related to the Al(111) surface [hereafter named as ($\sqrt{7}\times\sqrt{7}$) and ($3\times3$), respectively]. For the $\sqrt{7}\times\sqrt{7}$) structure, the scanning tunneling microscopy (STM) images display one bright protrusion in the unit cell[19-21, 25], and for the ($3\times3$) structure, most STM images show the same feature[19-21, 23, 29] and a few display two bright protrusions[8, 24]. However, it is difficult to determine the crystal structures from the experimental information, especially since a recent study has shown that even a simple honeycomb STM image can correspond to an unexpected complex structure[31]. Many structural models have been proposed for these two phases, for example, honeycomb (one of six[25] or eight[19] Ge atoms uplifted in the unit cell) lattices for the ($\sqrt{7}\times\sqrt{7}$) structure, and honeycomb (one[22] or two[8] of eight Ge atoms uplifted in the unit cell) and kagome[23] (one of ten Ge atoms uplifted in the unit cell) lattices for the (3 $\times$ 3) structure. Yet, Martinez et al.[18] questioned the germanene model for the (3 $\times$ 3) structure and proposed an alloy structure $Al_3Ge_5$, since they detected the presence of Al atoms within the formed surface layer by surface analysis techniques. Recently, by surface X-ray diffraction, a two-layer surface alloy[26] and a honeycomb alloy[56] were suggested for the (3 $\times$ 3) structure and the ($\sqrt{7} \times \sqrt{7}$) structure, respectively. Furthermore, density-functional theory (DFT) calculations indicated that the formation of the alloy is energetically favorable[27-28]. Overall, the exact atomic structure remains unclear, and there is no consensus even on the formation of germanene or surface alloys. Previous attempts have usually used empirical methods to modify the crystal structure to fit the experimental data, but this ad hoc approach often produces conflicting structural models especially when there are multiple types of experimental data[18, 22-23]. Recently, the global structural optimization algorithms have brought us a powerful tool to solve these complex structural problems, which has been successfully used in various systems[32]. In particular, the atomic-scale ultra-high-resolution STM images have been gained in a recent work[21], which can provide useful information to examine these structures obtained by the techniques of global optimizations.

In this work, we systematically explore the crystal structures that may form by Ge deposition onto Al(111) surface, by combining DFT and the global optimization algorithm USPEX[33-36]. When comparing the stability of structures, we consider various van der Waals density functionals in DFT and use high-level random-phase approximation (RPA)[37-38] to deal with conflicting structures. By constructing the convex hull diagram, we find three thermodynamically stable honeycomb alloy phases $Al_3Ge_3$/Al(111)($\sqrt{7} \times \sqrt{7}$), $Al_4Ge_2$/Al(111)($\sqrt{7} \times \sqrt{7}$), and $Al_3Ge_4$/Al(111)(3 $\times$ 3), of which the latter two phases are newly discovered. Note that a honeycomb alloy on Al(111)($\sqrt{7} \times \sqrt{7}$) was suggested in reference 56, but its exact structure is unknown because the full text has not been published yet, and we assume here that he proposed $Al_3Ge_3$/Al(111)($\sqrt{7} \times \sqrt{7}$). The simulated STM images of $Al_3Ge_3$/Al(111)($\sqrt{7} \times \sqrt{7}$) and $Al_3Ge_4$/Al(111)(3 $\times$ 3) are in excellent agreement with the experimental atomic-resolution STM images. In particular, the calculated quantitative low-energy electron diffraction (LEED) $I(V)$ curves of $Al_3Ge_4$/Al(111)(3×3) agree well with the experimental results with an overall Pendry reliability factor (Rp)[39] of 0.15. In addition, our results show that the formation of germanene on Al(111) is energetically unfavorable due to the positive formation energy calculated at the RPA level.

## Computational Method

The first-principles DFT calculations are carried out with the Perdew-Burke-Ernzerh generalized gradient approximation functional[40] and the VASP code[41-43]. The projected augmented wave method[44] is employed to describe the electron-ion interaction with $3s^2 3p^1$ and $3d^{10} 4s^2 4p^2$ treated as valence states for Al and Ge, respectively. The Al(111) surface is presented by a repeated slab with four Al layers, and the slabs are separated by a ~21 Å vacuum. The lattice constants of bulk Al and Ge are determined to be 4.038 and 5.759 Å, respectively. The cutoff energy for the plane-wave basis is set to 310 eV. For Al(111)($\sqrt{7}\times\sqrt{7}$) surface, a 7 × 7 × 1 and 11 × 11 × 1 k-point mesh is used for structural optimizations and static calculations, respectively. Similarly, for Al(111) (3×3) surface, a 6 × 6 × 1 and 9 × 9 × 1 k-point mesh is used. All atomic positions are fully relaxed with the Hellmann-Feynman force on each atom converged to ≤ 0.01 eV/Å expect the two bottom layers of the slab. We consider PBE-D2[45], PBE-D3[46], and optPBE-vdW[47-48] functionals to compare the stability of structures and use RPA[37-38] to resolve the conflicting results. For RPA calculations, we use 4 × 4 × 1 and 5 × 5 × 1 k-point mesh with a ~12 Å vacuum for Al(111) (3 × 3) and Al(111)($\sqrt{7}\times\sqrt{7}$) surface, respectively, and all these settings have been carefully checked to ensure that the relative stability of the structure is correctly described. The Tersoff-Hamann theory[49] is used to simulate STM images with 12 × 12 × 1 k-point mesh. The global optimization-based structure search is performed using the USPEX code with the *ab initio* evolutionary simulations[33-36].

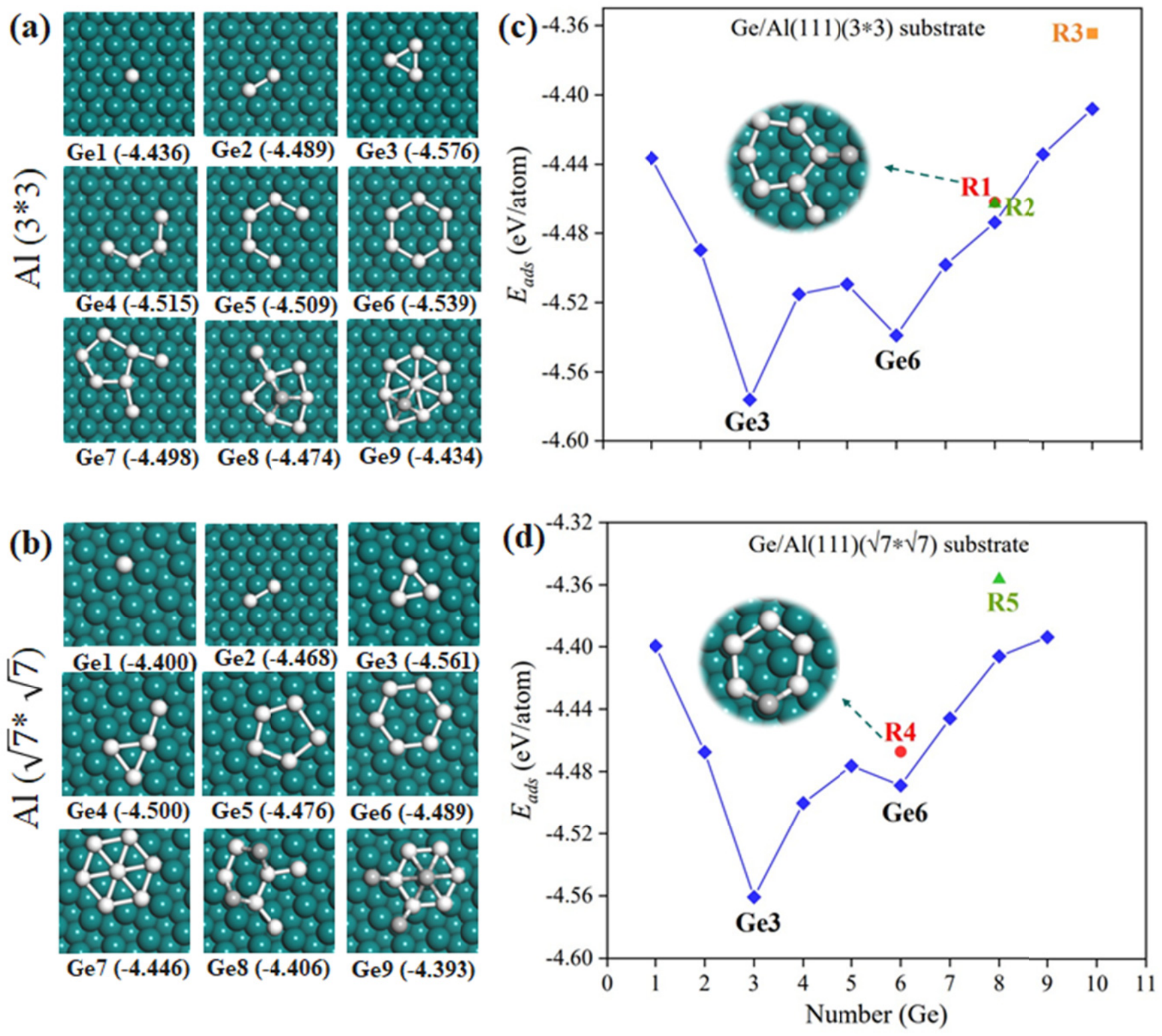

**Figure 1.** The lowest-energy structures are composed of pure Ge atoms: (a) on Al(3×3), (b) on Al($\sqrt{7}\times\sqrt{7}$). For clarity, we only show the structure within a unit cell on the surface. Comparison of the adsorption energies of structures with different numbers of Ge atoms: (c) on Al(3×3), (d) on Al($\sqrt{7}\times\sqrt{7}$). R1 to R5 represent the previously proposed structures, and the crystal structures are shown in Fig. S1c. The large (green) and small (white and grey) spheres represent Al and Ge atoms, respectively. Grey spheres represent uplifted Ge atoms.

## Results and Discussion

We first investigate the possible structures composed of pure germanium atoms on the Al(111) surface. Based on the experimentally observed periodicity, we carried out fix-composition global structure optimization on the Al(3×3) and Al($\sqrt{7}\times\sqrt{7}$) surfaces. One to ten Ge atoms and one to nine Ge atoms were considered for surfaces Al(3×3) and Al($\sqrt{7}\times\sqrt{7}$), respectively. A large number of structures were produced by the algorithm, whose defined adsorption energy was calculated as $E_{ads} = (E_{tot} - E_{sub}) / N_{Ge})$, where $E_{tot}$ is the energy of the whole system; $E_{sub}$ is that of the isolated substrate; $N_{Ge}$ denotes the number of Ge atoms on the surface. Figures 1a and 1b show the lowest-energy structures at each composition on the Al(3 × 3) and Al($\sqrt{7} \times \sqrt{7}$) surfaces, respectively. The corresponding adsorption energies are shown in Fig. 1c and 1d, which also include that of the previously proposed structures (labeled by R1 to R5). Our results show that the energetically most favorable structure is a triangular island formed by three Ge atoms for both surfaces [named $Ge_3$/Al(3 × 3) or $Ge_3$/Al($\sqrt{7} \times \sqrt{7}$); this definition also applies to structures with other numbers of Ge atoms], where each Ge atom is located at the FCC (face-centered cubic) site of the substrate. The adsorption energy of $Ge_3$/Al(3 × 3) is slightly lower than that of $Ge_3$/Al($\sqrt{7} \times \sqrt{7}$), indicating repulsive interactions between the $Ge_3$ triangular islands. To confirm whether this structure is experimentally observed, we simulated its STM image. As shown in Fig. S1a, three bright protrusions are displayed in a unit cell, which is distinctly different from the experimental results[8, 19, 21] and is therefore excluded.

Germanene with buckled honeycomb lattice has attracted much attention, due to its predicted exotic physical properties[2-4]. Our results show that the lowest-energy germanene structures on both surfaces are in a flat honeycomb configuration (named $Ge_6$), as shown in Fig. 1a and 1b. However, its adsorption energy is higher than that of the $Ge_3$ triangular islands structure, and the simulated STM image shows six bright protrusions in a unit cell (see Fig. S1b) that is different from the experimental observations[8, 19, 21]. Several buckled honeycomb germanene models have been proposed, such as one[22] or two[8] of eight Ge atoms uplifted in the unit cell for Al(3 × 3) surface (here named as R1 or R2) and one of six[25] or eight[19] Ge atoms uplifted in the unit cell for Al($\sqrt{7} \times \sqrt{7}$) surface (here named as R4 or R5). As shown in Fig. 1c and 1d, the adsorption energies of these structures are higher than that of the flat honeycomb $Ge_6$, and they are not even the lowest-energy structures of the corresponding Ge compositions. The previously proposed kagome lattice with ten Ge atoms[23] for Al(3 × 3) surface also has much higher energy (see R3 of Fig. 1c).

*Table 1.*Calculated Formation Energies(eV/atom) of $Ge_3$ and $Ge_6$ with different density functionals

| | Ge3 | Ge6 |
|---|---|---|
| PBE | -0.058 | -0.021 |
| PBE-D2 | 0.001 | 0.002 |
| PBE-D3 | -0.088 | -0.072 |
| optPBE-vdW | -0.007 | 0.033 |
| RPA | 0.0623 | 0.114 |

It is interesting to revisit whether Al(111) surface is a promising substrate for growing germanene, as the above low-energy structures, such as $Ge_3$, have not been observed in the experiments. To evaluate the long-range van der Waals (vdW) interaction (the energies in Fig.1 are calculated at the PBE level), we have considered PBE-D2, PBE-D3, and optPBE-vdW functionals. Although they show almost the same trend in the relative stability of the structures, the exact formation energy varies considerably (see Fig. S2). The formation energies of the lowest-energy triangular island $Ge_3$/Al(3×3) and flat honeycomb $Ge_6$/Al(3×3) calculated by different functionals are listed in Table 1, which would produce different conclusions about the growth of these two structures. According to the results of PBE or PBE-D3, both phases can be grown experimentally by adjusting the chemical potential of Ge, while only $Ge_3$/Al(3×3) can be grown according to that of optPBE-vdW, and no phases can be grown according to that of PBE-D2. To resolve this conflict, we use a high-level RPA that can accurately describe the van der Waals interactions. Our calculations show that the formation energies of both structures are positive, which are 0.063 eV/atom and 0.114 eV/atom for $Ge_3$/Al(3×3) and $Ge_6$/Al(3×3), respectively (see Table 1). This explains why these phases consisting of Ge atoms, such as $Ge_3$ or $Ge_6$, were not observed experimentally. Therefore, the growth of germanene on the Al(111) surface is energetically unfavorable, and it tends to detach from the surface to form three-dimensional Ge crystallites.

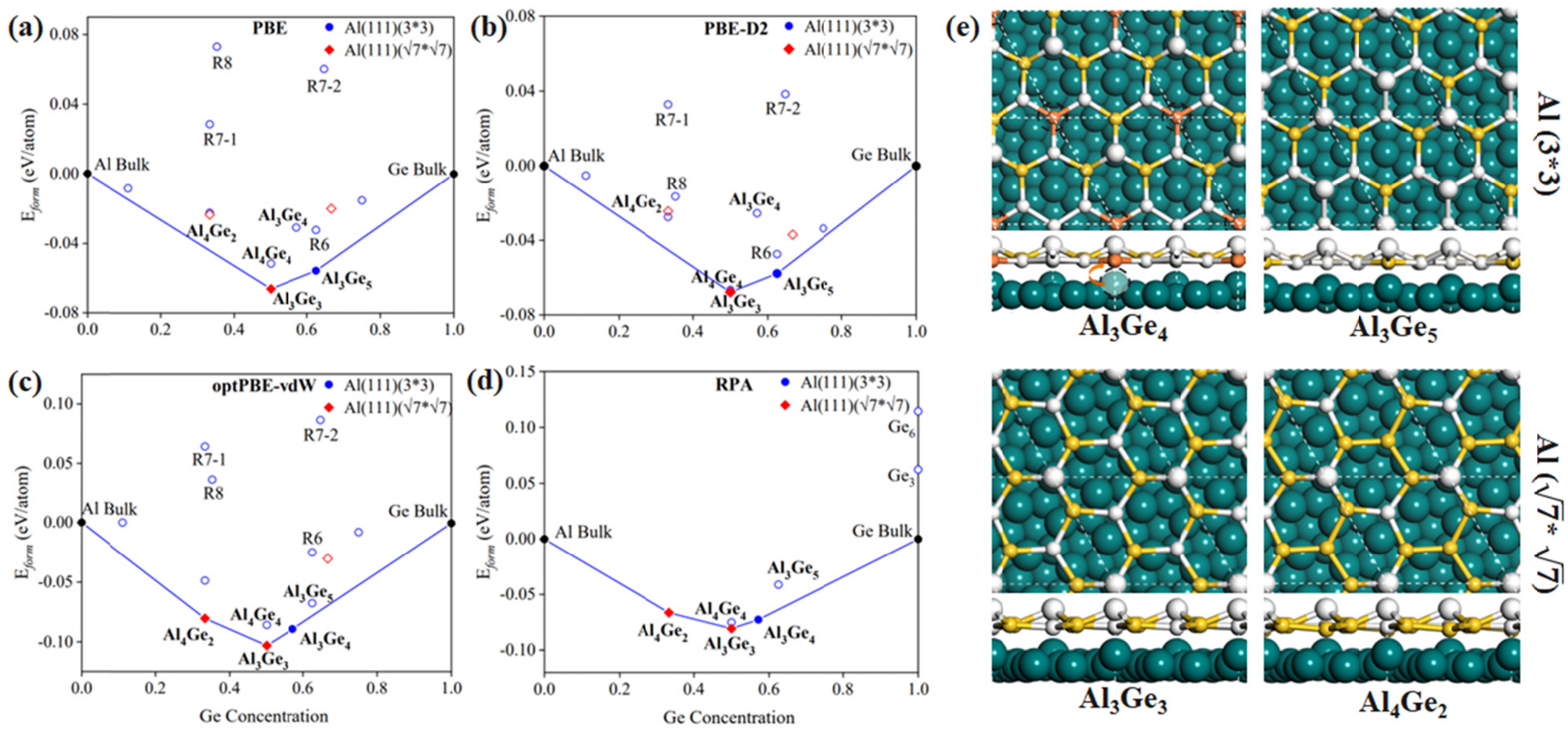

**Figure 2.** Thermodynamic convex hulls of the Al-Ge surface alloys were calculated with different density functionals in DFT: (a) PBE, (b) PBE-D2, (c) optPBE-vdW, and (d) RPA. Solid circles and rhombus denote stable phases. R6 to R8 represent the previously proposed structures (R6: $Al_3Ge_5$; R7-1: $Al_2Ge$; R7-2: $Ge_8/Al_2Ge$; R8: $Al_4Ge_4/Al_7Ge_2$), and the crystal structures are shown in Fig. S4. (e) The thermodynamic stable phases $Al_3Ge_4$/Al(3×3), $Al_3Ge_3$/Al($\sqrt{7}\times\sqrt{7}$), $Al_4Ge_2$/Al($\sqrt{7}\times\sqrt{7}$), and meta-stable phase $Al_3Ge_5$/Al(3×3). The green spheres represent Al atoms in the substrate; yellow spheres represent Al atoms in the alloy layer; white spheres represent Ge atoms; big white spheres represent the uplifted Ge atoms. The orange spheres in $Al_3Ge_4$/Al(3×3) represent the Al atoms coming from the substrate and the black dotted circle represents the vacancy left in the substrate.

We now use the variable-composition evolutionary simulations to investigate the possible alloy structures on Al(3×3) and Al($\sqrt{7}\times\sqrt{7}$) surfaces. Single-layer and two-layer alloys of various compositions were systematically explored in our calculations. To compare the relative stability of the alloys with different compositions, we constructed the convex hull diagram[32] by calculating the energies of formation as $E_{form} = (E_{tot} - E_{sub} - x\times E_{Al} - y\times E_{Ge})/(x+y)$, where $E_{tot}$ is the energy of the whole system; $E_{sub}$ is that of the isolated substrate; $E_{Al}$ and $E_{Ge}$ represent the energy per Al atom and Ge atom in the bulk; x and y denote the number of Al atoms and Ge atoms of alloy $Al_xGe_y$ on the surface. A structure whose formation energy lies on the convex hull is thermodynamically stable. In Fig. 2a, we show the formation energy (only negative values) of the most stable structure at each composition for both surfaces, and that of the previously proposed alloy structures are also shown (labeled by R6 to R8). The energies in Fig. 2a were calculated at the PBE level, and we have used PBE-D2, PBE-D3, and optPBE-vdW functionals to evaluate the vdW interaction, as shown in Fig. 2b, Fig. S3a, and Fig. 2c, respectively. The PBE, PBE-D2, and PBE-D3 functionals show the same results that the $Al_3Ge_3/Al(\sqrt{7}\times\sqrt{7})$ and $Al_3Ge_5/Al(3\times3)$ are the stable phases, however, the optPBE-vdW functional shows the $Al_3Ge_3/Al(\sqrt{7}\times\sqrt{7})$, $Al_2Ge_4/Al(\sqrt{7}\times\sqrt{7})$ and $Al_3Ge_4/Al(3\times3)$ are the stable phases. Since there are only a few stable structures involved (SCAN meta-GGA[50-51] has also been used to check these results as shown in Fig. S3b), we can use high-level RPA to resolve this conflict, which can accurately describe the van der Waals interactions. The formation energies of the above low-energy structures $Al_3Ge_3/Al(\sqrt{7}\times\sqrt{7})$, $Al_4Ge_2/Al(\sqrt{7}\times\sqrt{7})$, $Al_4Ge_4/Al(3\times3)$, $Al_3Ge_4/Al(3\times3)$, and $Al_3Ge_5/Al(3\times3)$ are calculated at the RPA level (see Fig. 2d). Interestingly, the opt-VDW functional shows consistent results with the RPA, indicating that it is suitable for this Ge/Al system. Three thermodynamically stable structures $Al_3Ge_3/Al$ $(\sqrt{7}\times\sqrt{7})$, $Al_4Ge_2/Al$ $(\sqrt{7}\times\sqrt{7})$, and $Al_3Ge_4/Al(3\times3)$ is finally determined, whose detailed crystal structures are shown in Fig. 2e. They are all bulked honeycomb configures with one Ge atom unfilled in a unit cell. The $Al_3Ge_4/Al(3\times3)$ is interesting, in which an Al atom of the substrate is drawn to the alloy layer to construct the honeycomb lattice, remaining a vacancy in the substrate. It is worth noting that the $Al_3Ge_3/Al(\sqrt{7}\times\sqrt{7})$ was reported in very recent work by surface X-ray diffraction[56], indicating the reliability of our method. Other structures proposed previously have much higher formation energies (see Fig 2a, 2b, and 2c), such as single-layer alloys $Al_3Ge_5$[18] (here named R6) and $Al_2Ge$[27] (here named R7-1), two-layer alloys $Ge_8/Al_2Ge$[27] (the top layer is $Ge_8$ and the sub-layer is $Al_2Ge$; here named R7-2) and $Al_4Ge_4/Al_7Ge_2$[26] (the top layer is $Al_4Ge_4$ and the sub-layer is $Al_7Ge_2$; here named R8).

To compare with the experimentally observed structures, we simulated STM images of the three stable alloy phases. We strictly use the experimental bias to simulate the STM, that is, -1.2 V and -0.03 V for the structures on $Al(3\times3)$ and $Al(\sqrt{7}\times\sqrt{7})$, respectively[21]. Remarkably, the simulated STM images of $Al_3Ge_4$/Al(3×3) and $Al_3Ge_3$/Al($\sqrt{7}\times\sqrt{7}$) are in perfect agreement with the experimental atomic-resolution STM images (see Fig. 3). For the experimental STM image of the structure on the Al(3×3) surface, an obvious feature was the observed dark clover pattern with a light-colored bright spot at the center (see Fig. 3a). This clover pattern in our $Al_3Ge_4$/Al(3×3) structure originates from the vacancy within the Al substrate. If we fill this vacancy with additional Al atoms, the $Al_4Ge_4$/Al(3×3) structure is obtained, whose formation energy is higher than that of the $Al_3Ge_3$/Al($\sqrt{7}\times\sqrt{7}$) at the same composition (see Fig. 2d). The simulated STM image of $Al_4Ge_4$/Al(3×3) displays a distinct bright spot in the center of the clover (see Fig. S5a), which is different from the experimental result. For the structure on Al($\sqrt{7}$ $\times\sqrt{7}$) surface, the experimental STM image shows a bright spot surrounded by three identical halos (see Fig. 3c), which can well be reproduced by the simulated STM image of the $Al_3Ge_3$/Al($\sqrt{7}\times\sqrt{7}$) (see Fig. 3d). We also simulated the STM image of the $Al_4Ge_2$/Al($\sqrt{7}\times\sqrt{7}$), which does not conform as well to the experimental results as that of the $Al_3Ge_3$/Al($\sqrt{7}\times\sqrt{7}$) (see Fig. S5b). It should be pointed out that a few works have reported the STM with two bright spots in a unit cell[8, 24, 30], which may correspond to the meta-stable $Al_3Ge_5$/Al(3×3) structure with two Ge atoms uplifted in a unit cell (the formation energy, the structure, and the simulated STM image are shown in Fig. 2d, 2e and Fig. S5d, respectively). In particular, the $Al_3Ge_5$/Al(3×3) can be transformed into a structure with an uplifted Ge atom with only 0.015 eV energy higher (see Fig. S5c), which may explain the experimental tip-induced switching between structures with one bright spot and two bright spots in the STM image[30].

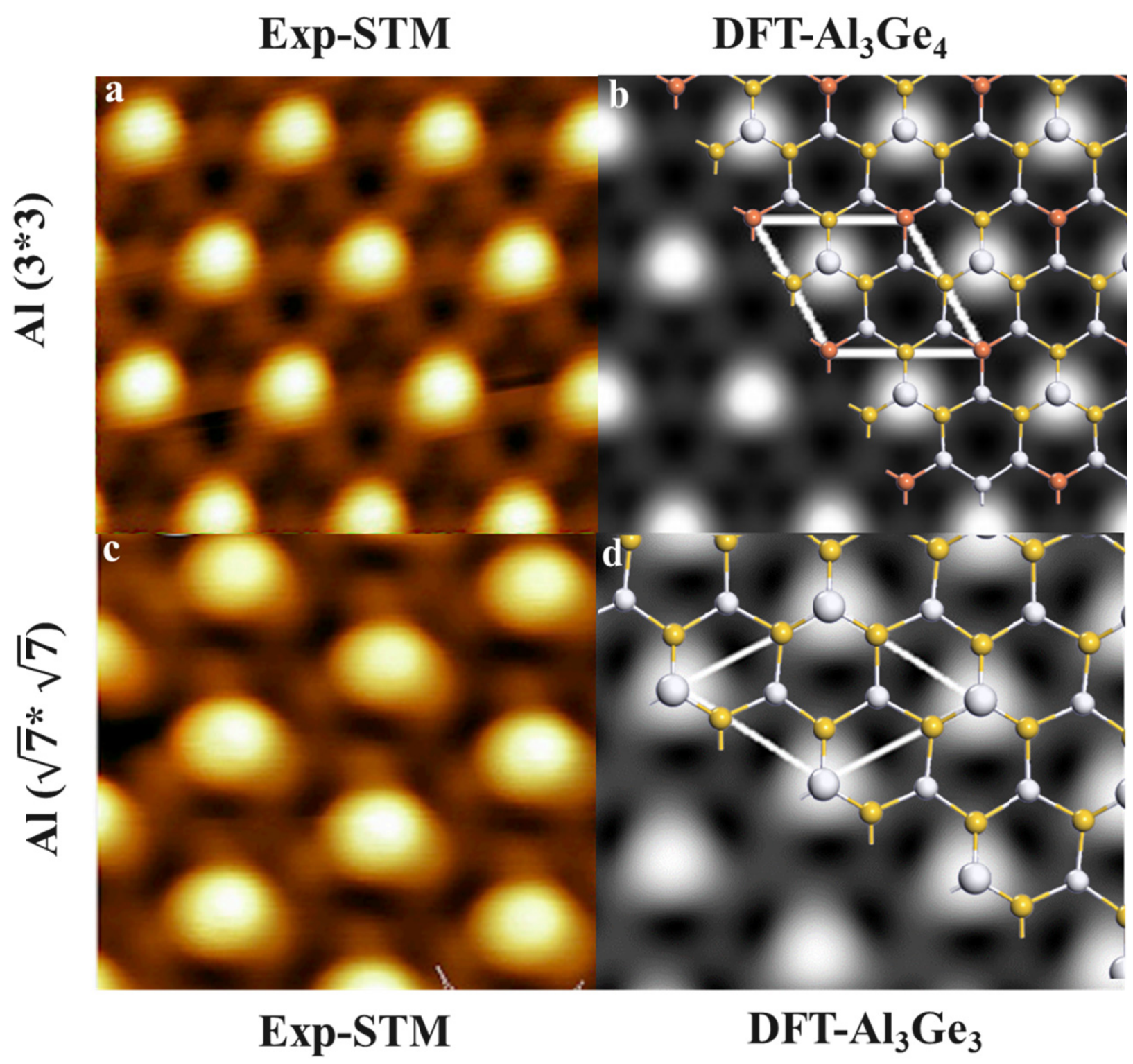

**Figure 3.** The experimental STM images: (a) on Al(3×3), (c) on Al($\sqrt{7}\times\sqrt{7}$). The simulated STM images: (b) for $Al_3Ge_4$/Al(3×3), (d) for $Al_3Ge_3$/Al($\sqrt{7}\times\sqrt{7}$). The simulation uses constant current. Reproduced with permission[21].

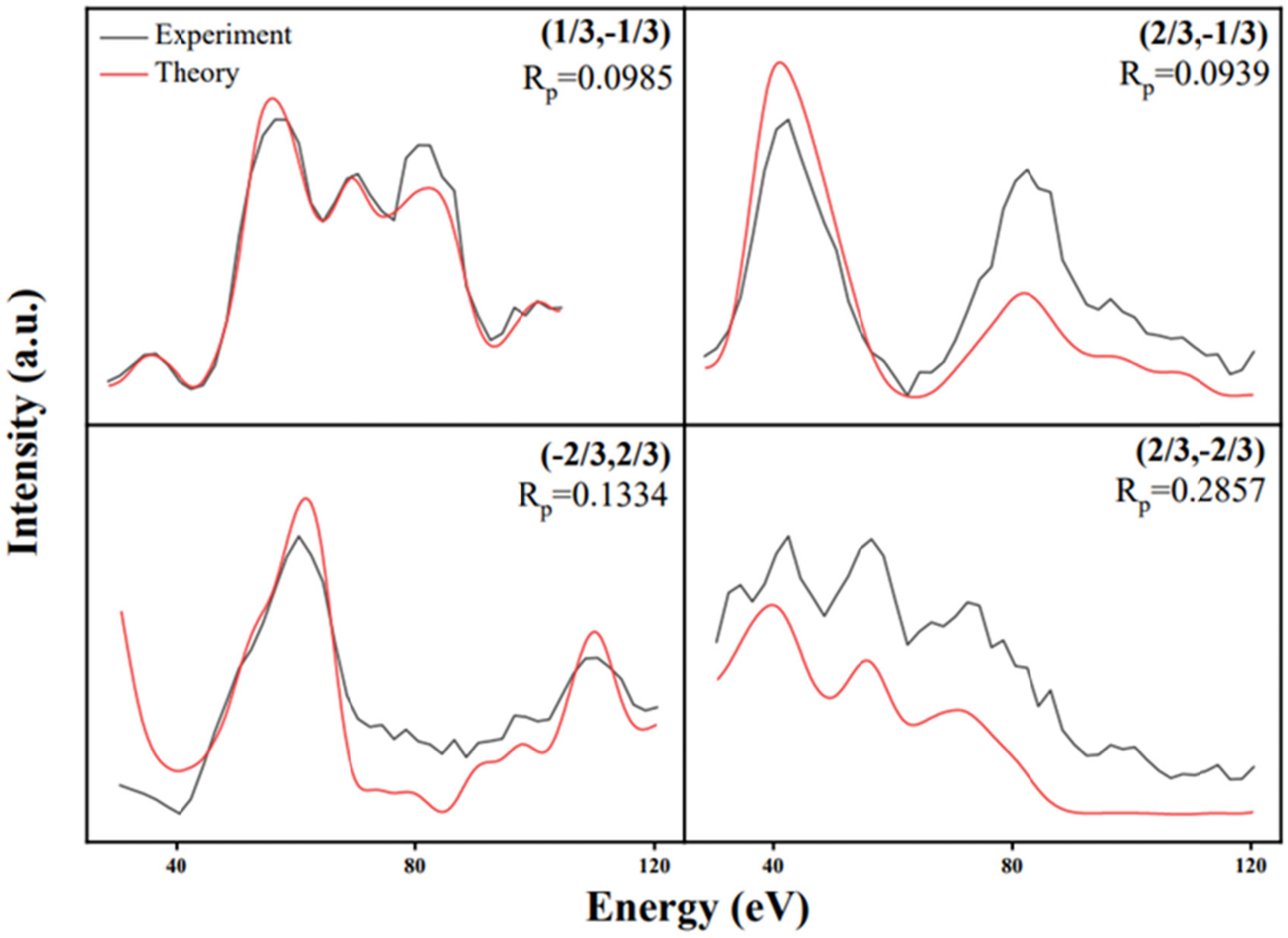

**Figure 4.** Comparison between experimental and simulated LEED $I(V)$ curves for the $Al_3Ge_4$/Al(3×3) structure. Reproduced with permission[18].

Quantitative analyses of LEED *I*(*V*) curves can provide very useful information for identifying surface structures[52]. Since the LEED *I*(*V*) data of the structure on Al(3×3) is available[18], we carried out a dynamic calculation of the LEED *I*(*V*) curves for $Al_3Ge_4$/Al(3×3). The calculation is performed using the AQuaLEED package[53-54], which is based on the Barbieri/Van Hove SATLEED package[52, 55]. The crystal structure of $Al_3Ge_4$/Al(3 × 3) optimized from DFT calculations is used as the starting configuration, and then quantitative LEED was used to refine the atomic positions. The atomic coordinates derived from quantitative LEED are consistent with that from the DFT calculation, with a maximum displacement difference of 0.34 Å. The comparison between calculated and measured *I(V)* curves is shown in Fig. 4, with an overall Pendry reliability factor (Rp)[39] of 0.153. Usually, an R-factor of about 0.2 is considered a good fit between experiment and theory[39]. This indicates that the calculated LEED *I*(*V*) curves of the $Al_3Ge_4$/Al(3×3) are excellent agreeing with experimental results.

**Conclusion**

In summary, we have systematically investigated the possible structures formed by Ge deposition onto Al(111) surface using the state-of-the-art global optimization algorithm USPEX. Different van der Waals density functionals and high-level RPA were used to determine the stable structure, where optPBE-vdW exhibited consistent results with the RPA. Three thermodynamically stable honeycomb alloy phases $Al_3Ge_3$/Al($\sqrt{7}\times\sqrt{7}$), $Al_4Ge_2$/Al($\sqrt{7}\times\sqrt{7}$), and $Al_3Ge_4$/Al(3×3) are identified, of which the latter two phases are newly discovered. Our results show that the two debated experimental phases are honeycomb alloys $Al_3Ge_3$/Al($\sqrt{7}\times\sqrt{7}$) and $Al_3Ge_4$/Al(3×3), which can well reproduce the experimental atomic-resolution STM images. In particular, the calculated LEED *I*(*V*) curves of $Al_3Ge_4$/Al(3 × 3) agree well with the experimental results. $Al_3Ge_4$/Al(111)(3×3) is an interesting structure with a vacancy in the substrate, which accounts for the dark clover pattern in the experimental STM image. Furthermore, our results show that the growth of germanene on Al(111) is energetically unfavorable with positive formation energy calculated at the RPA level. This work indicates that the fabrication of germanene may still be a challenge.

**Corresponding Author**

*E-mail: xuh@sustech.edu.cn

**ACKNOWLEDGMENT**

This work is supported by the National Natural Science Foundation of China (No. 11974160), the Science, Technology, and Innovation Commission of Shenzhen Municipality (Nos. RCYX20200714114523069), and the Center for Computational Science and Engineering at Southern University of Science and Technology.

# Supporting Information for: Identifying the Alloy Structures of Germanene Grown on Al(111) Surface

Feini Yan,[a] Shaogang Xu,[a] Chao He,[a] Changchun He,[a] Changming Zhao,[a] and Hu Xu[a,*]

[a]Department of Physics, Southern University of Science and Technology, Shenzhen518055, China

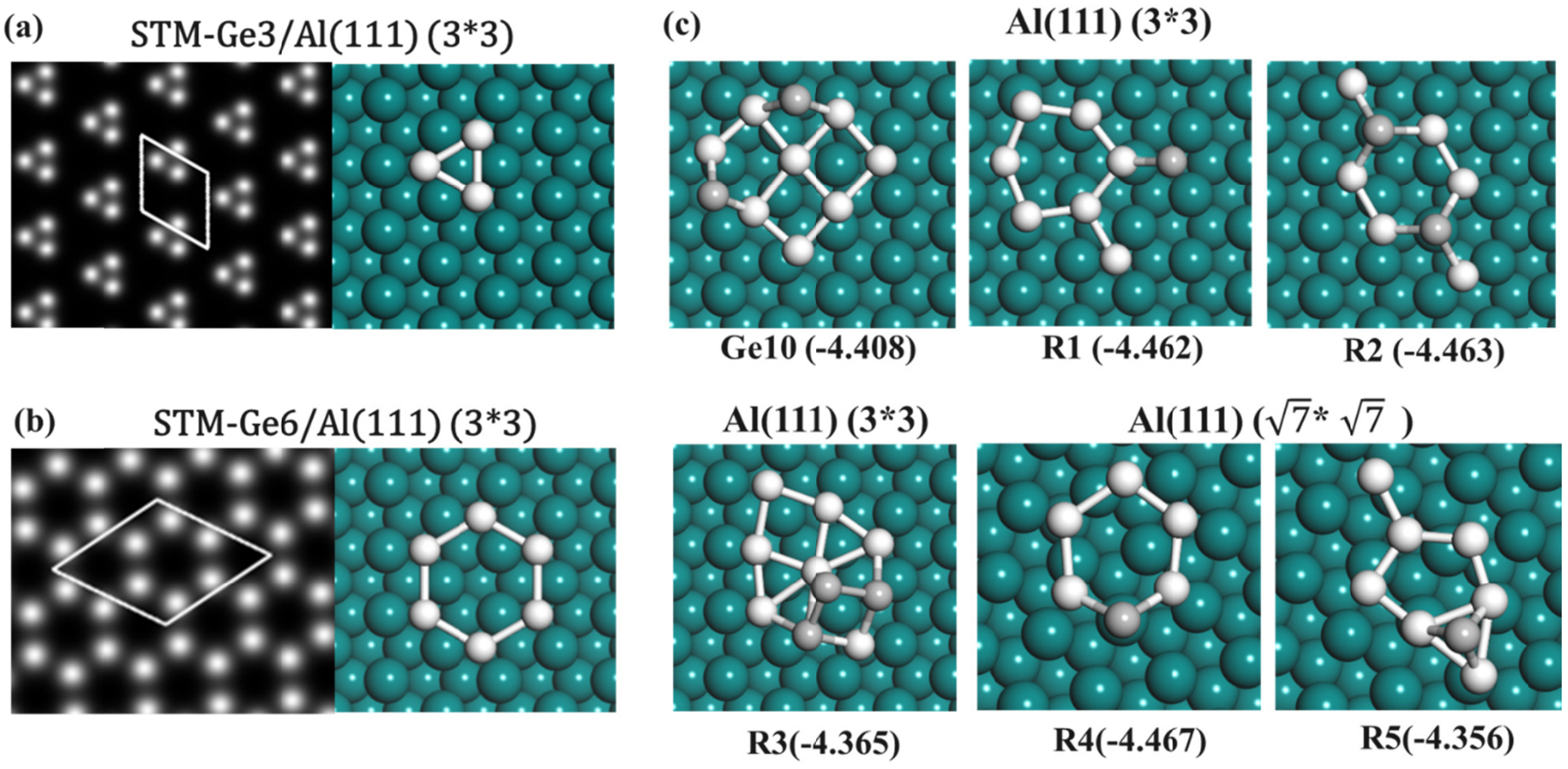

Figure S1: (a) Simulated STM of $Ge_3$/Al(3×3). (b) Simulated STM of $Ge_3$/Al(3×3). The simulations were performed at a bias voltage of -1.2 V with constant height. (c) $Ge_{10}$ is the structure searched by the global optimization. R1 to R5 are the previously proposed structures.

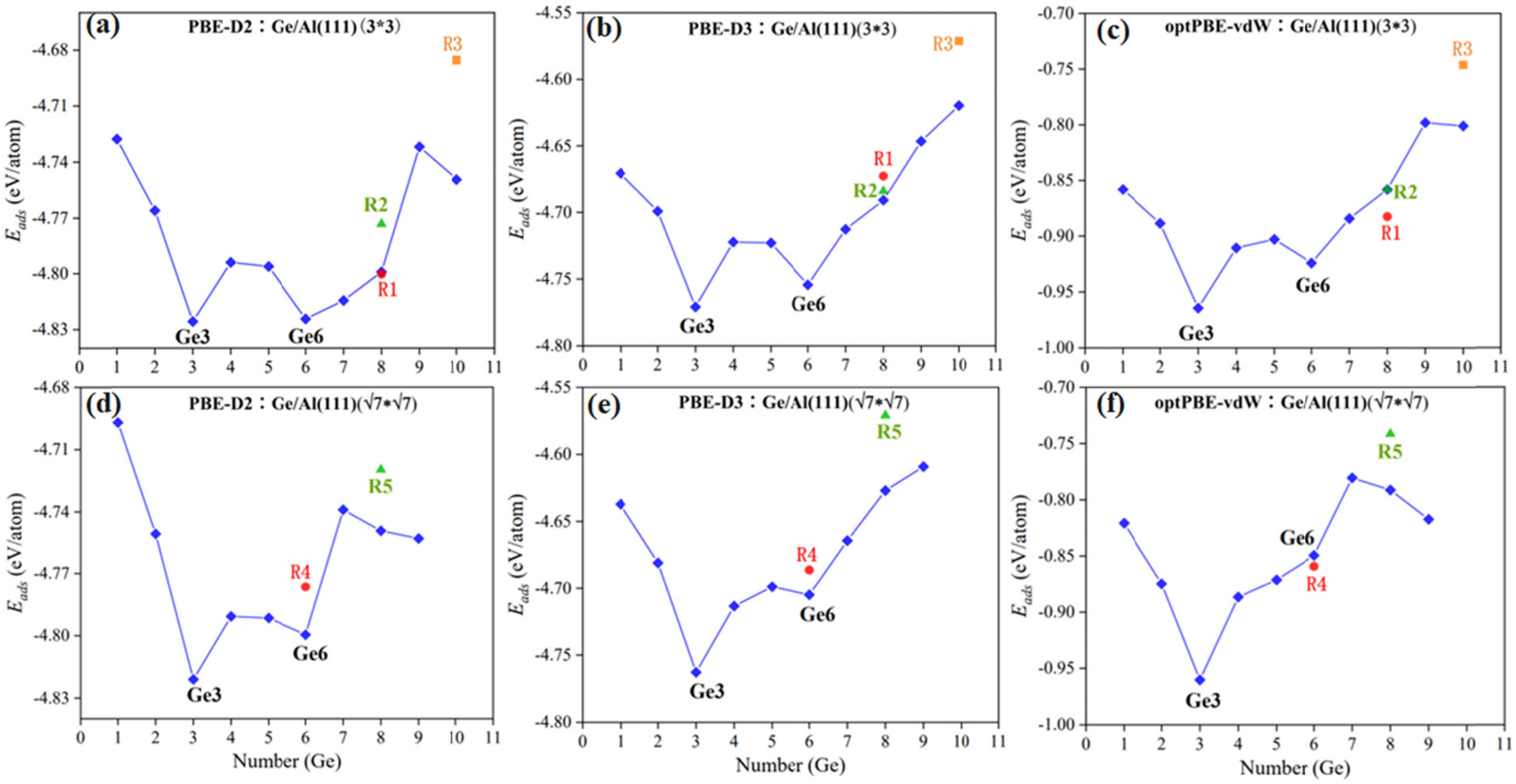

Figure S2: Comparison of the adsorption energies of structures with different numbers of Ge atoms: (a)-(c) on Al(3×3), (d)-(f) on Al(√7×√7). Different van der Waals density functionals are used: (a) and (b): PBE-D2; (b) and (e): PBE-D3; (c) and (f): optPBE-vdW.

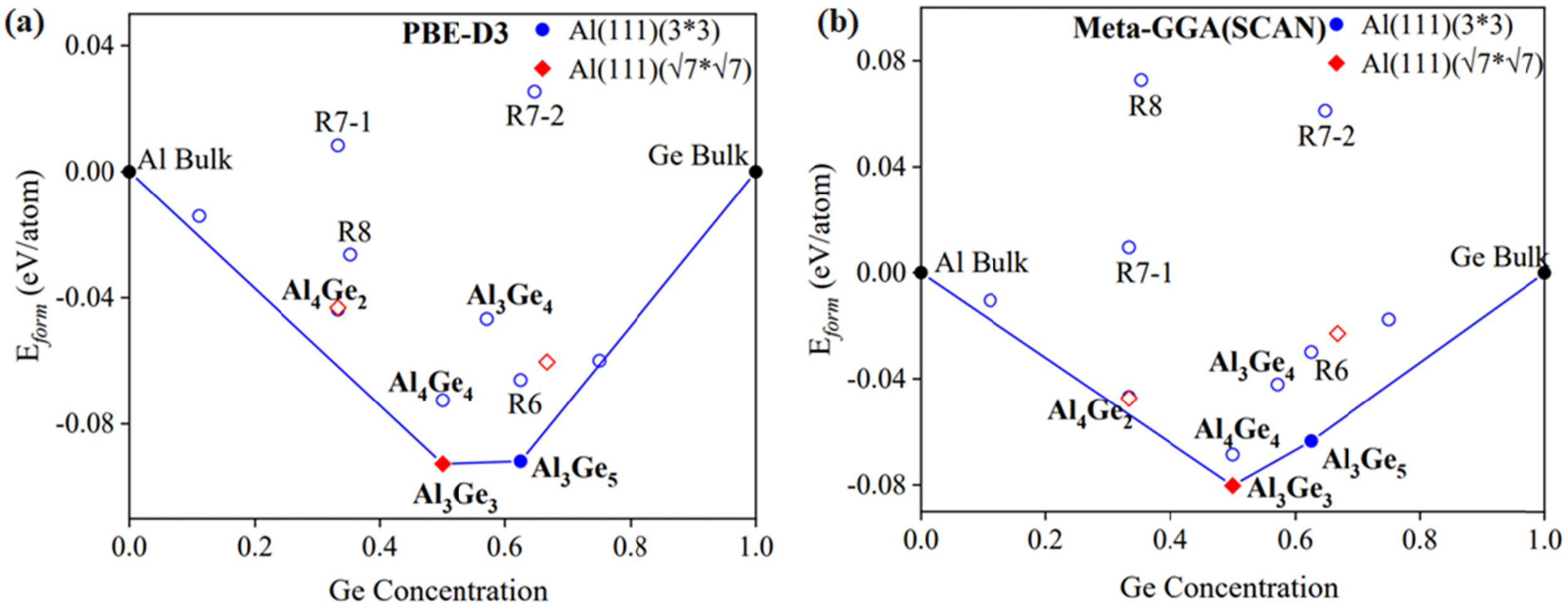

Figure S3: Thermodynamic convex hulls of the Al-Ge surface alloys calculated with different density functionals in DFT: (a) PBE-D3 and (b) Meta-GGA (SCAN).

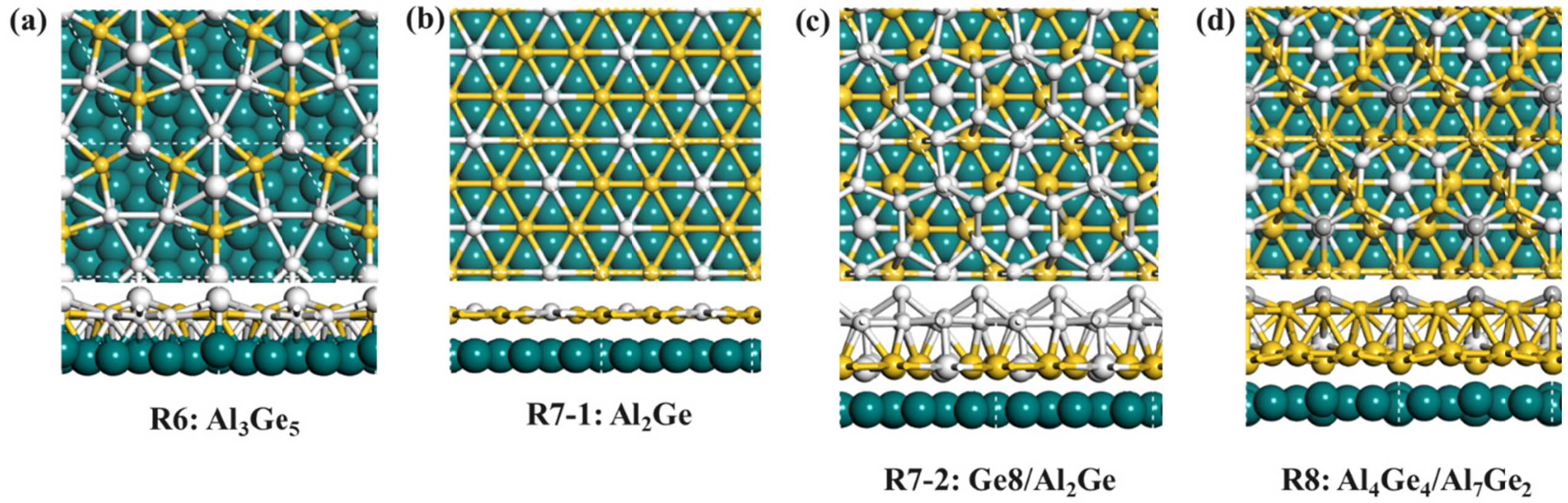

Figure S4: The the previously proposed alloy structure on Al(111)(3*3). (a) R6: $Al_3Ge_5$; (b) R7-1：$Al_2Ge$; (c) Ge8/$Al_2Ge$; (d) $Al_4Ge_4$/$Al_7Ge_2$.

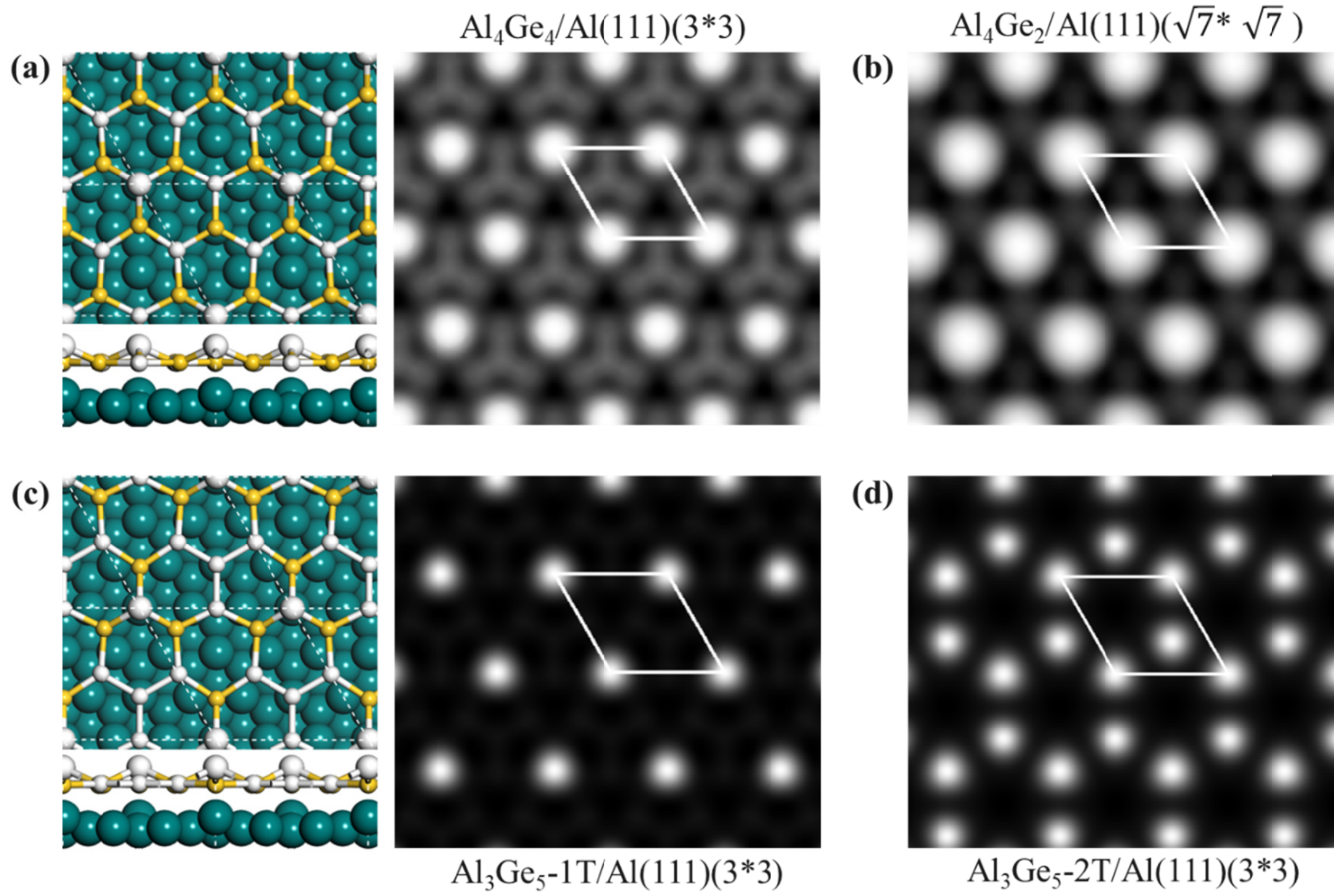

Figure S5: Simulated STM images and crystal structures of different composition alloys: (a) $Al_4Ge_4$/Al(111)(3*3) (U= -1.2eV, constant current); (b) $Al_4Ge_2$/Al(111)($\sqrt{7} * \sqrt{7}$) (U= -0.03eV, constant current); (c) $Al_3Ge_5$-1T/Al(111)(3*3) (A Ge atom protrudes upwards) (U= -1.2eV, constant height); (d) $Al_3Ge_5$-/Al(111)(3*3) (Two Ge atoms protrude upwards) (U= -1.2eV, constant height).